\documentclass[aip,jcp,reprint,groupedaddress]{revtex4-1}
\usepackage{graphicx}
\usepackage{color}
\usepackage{amsmath}
\usepackage{tabularx}

\newcommand{\vect}[1]{\boldsymbol{#1}}

\begin{document}

\title{
  Aggregation of theta-polymers in spherical confinement
}
\author{Johannes Zierenberg}
\email[]{zierenberg@itp.uni-leipzig.de}
\author{Marco Mueller}
\email[]{mueller@itp.uni-leipzig.de}
\author{Philipp Schierz}
\email[]{schierz@itp.uni-leipzig.de}
\author{Martin Marenz}
\email[]{marenz@itp.uni-leipzig.de}
\author{Wolfhard Janke}
\email[]{janke@itp.uni-leipzig.de}
\affiliation{
  Institut f\"ur Theoretische Physik, 
  Universit\"at Leipzig, 
  Postfach 100\,920,
  04009 Leipzig, Germany
}

\date{\today}

\begin{abstract}
  We investigate the aggregation transition of theta polymers in spherical
  confinement with multicanonical simulations. This allows for a systematic
  study of the effect of density on the aggregation transition temperature for
  up to $24$ monodisperse polymers.
  {Our results for solutions in the dilute regime show that 
  polymers can be considered isolated for all temperatures larger than the
  aggregation temperature, which is shown to be a function of the density.
  The resulting competition between single-polymer collapse and aggregation
  yields the lower temperature bound of the isolated chain approximation.
  We provide entropic and energetic arguments to describe the density
  dependence and finite-size effects of the aggregation transition for
  monodisperse solutions in finite systems.}
  This allows {us} to estimate the aggregation transition
  temperature of dilute systems in a spherical cavity, using a few simulations
  of small, sufficiently dilute polymer systems.
\end{abstract}


\maketitle


\section{Introduction}
Polymers are often studied as isolated chains. In nature, however, a system of
polymers or proteins is usually subject to geometrical confinement, {e.g. porous
media~\cite{Dullien1979}} or molecular crowding~\cite{Minton2000,Gershenson2010}.
This has an effect on structural as well as dynamic
properties.~\cite{Eggers2000,Schneider2011} It may also be the reason for
entropic forces in segregation processes.~\cite{Jun2010,Jung2012} 
The process of aggregation itself plays a role in biological systems as well as
in technological applications and material design, e.g., in the context of
photovoltaic cells.~\cite{Kim2011}

The effect of spherical confinement on the linear extension of polymers in
good solvent (modeled by a self-avoiding chain) has been studied by scaling
arguments~\cite{Sakaue2006}, Monte Carlo simulations~\cite{Cacciuto2006} and
Molecular Dynamics simulations~\cite{Arnold2007}. A relation between the free
energy of a single polymer and semi-dilute solutions was established. 
Adding short-range attraction to the excluded volume leads to a theta polymer
that exhibits a collapse transition from an extended coil at large temperatures
to a compact globule at lower temperatures. The effect of spherical confinement
on a single flexible theta polymer~\cite{Marenz2012} was shown to be different
to (rather stiff)
protein~\cite{Klimov2002,Friedel2003,Takagi2003,Rathore2006,Mittal2008} models.
For both cases, it was shown that the confinement shifts the location of the
collapse transition temperature.

Moreover, spherical confinement provides a safe basis for the study of density
effects in finite systems. For a steric confinement, there is at most an
effective repulsive interaction. In contrast to periodic boundary conditions it
allows to decrease or increase the density systematically without the
possibility that the aggregate, or even single polymers, may interact with
themselves across the boundaries. The influence of density on the aggregation
transition of two lattice proteins has been noticed recently~\cite{Ni2013} to be
similar to that on an ideal gas.
In this study we will extend this observation by investigating an off-lattice
polymer model that has been successfully applied to
peptide~\cite{JunghansPeptide} and polymer
aggregation~\cite{JunghansPolymer,Zierenberg2014} before.
{Focusing on the case of flexible homopolymers, we} will provide
entropic arguments for the density dependence and energetic arguments for
finite-size effects which leads to a reasonable description of the density
dependence in a spherical confinement.  {Among others, we
demonstrate the competition between single-chain collapse and multi-chain
aggregation in the dilute regime, showing the dominance of aggregation and the
consequences on structural properties of a single polymer.}

The paper is organized as follows: In Sec.~\ref{sec:model}, we briefly describe
the employed aggregation model together with the multicanonical method. We
mention the applied optimizations and relevant parameters.
Section~\ref{sec:results} contains all main results including a discussion of
the canonical picture, entropic arguments in the microcanonical picture, and a
description of finite-size effects. We finish with our conclusions on the effect
of density on the aggregation transition temperature in spherically confined
finite polymer systems in Sec.~\ref{sec:conclusion}.

%
%
\section{Model and Method} 
\label{sec:model}
\begin{figure}
  \includegraphics[width=6cm]{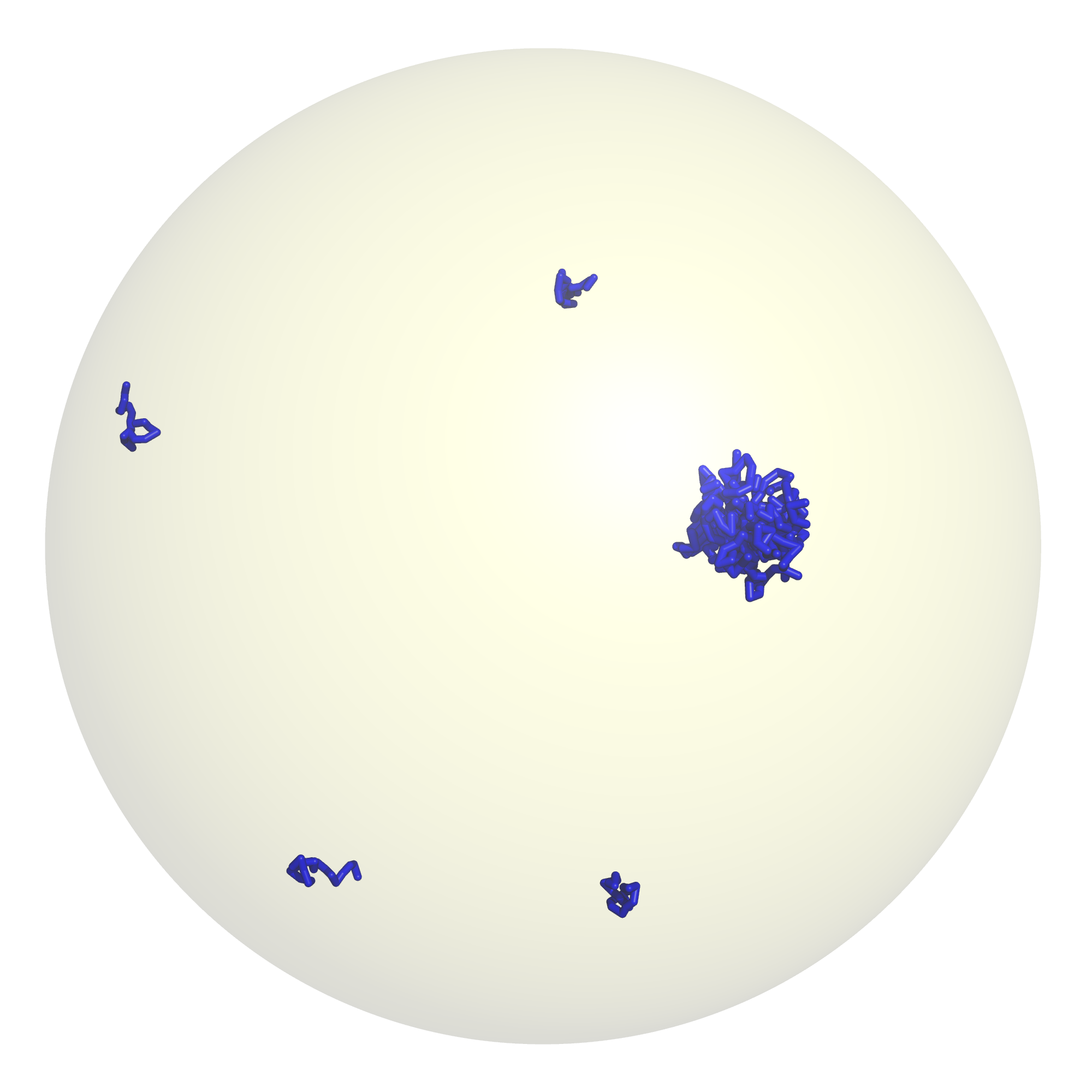}
  \caption{
    {
    Snapshot of a polymer system with $M=20$ polymers of length $N=20$ in a
    sphere with radius $R_S=30$. The snapshot was taken in the final
    production run of the multicanonical simulation and the conformation
    corresponds to an intermediate state inside the aggregation transition.
  }
  \label{fig:snapshot}
  }
\end{figure}

We consider a set of $M$ bead-spring polymers confined in a spherical cavity of
radius $R_S$. Each homopolymer consists of $N$ equal monomers aligned linearly
with a finitely extensible nonlinear elastic (FENE) potential between bonded
monomers and Lennard-Jones interaction between non-bonded monomers. The
interactions are parameterized as in
Refs.~\onlinecite{Binder2001,Schnabel2009,Zierenberg2014}, namely the FENE
potential
\begin{equation}
  V_{\rm FENE}(r)=-\frac{K}{2}R^2\ln\left(1-[(r-r_0)/R]^2\right),
\end{equation}
with $r_0=0.7$, $R=0.3$, and $K=40$, as well as the 12-6 Lennard-Jones potential
\begin{equation}
  V_{\rm LJ}(r) = 4 \epsilon \left[ (\sigma/r)^{12} - (\sigma/r)^6 \right],
\end{equation}
with $\epsilon=1$ and $\sigma=r_0/2^{1/6}$. For numerical reasons and in order
to be consistent with the aforementioned literature, the Lennard-Jones
potential is cutoff at $r_c=2.5\sigma$ such that
\begin{equation}
  V^*_{\rm LJ}(r) = 
  \begin{cases}
    V_{\rm LJ}(r)-V_{\rm LJ}(r_c) & r < r_c \\
    0                             & \mathrm{else}
  \end{cases}
\end{equation}
has the same qualitative behavior and is still continuous at $r_c$.  The
Lennard-Jones potential accounts for excluded volume and short-range attraction
such that each polymer can undergo a collapse transition.  There is no
distinction between the interaction of monomers within the same polymer or
between different polymers.  In general, we focus in this study on flexible
polymers with one exception, when we discuss the direct influence of the
density on a specific example of stiff polymers. Stiffness is introduced as a
penalty from the discretized polymer curvature. This results in a bending
potential
\begin{equation} 
  V_{\rm bend}(\theta) = \kappa\left(1-\cos\theta\right), 
\end{equation}
where $\theta$ is the angle between consecutive bond vectors. The polymer
system is constraint in a steric sphere such that conformations exceeding the
spherical confinement are forbidden{, for a snapshot see
Fig.~\ref{fig:snapshot}}.
{In general the radius of the sphere is much larger than the
  linear extension of a single polymer, which can be estimated within a
  self-avoiding walk picture to have a radius of gyration $R_{\rm gyr}\propto N^\nu$
  ($\nu\approx0.588$).~\cite{Clisby2010}
}

We employ Markov chain Monte Carlo simulations in the multicanonical
ensemble~\cite{MUCA,Janke1998} because aggregation shows characteristics of a
first-order phase transition for which this method was proven to be
particularly efficient. The method allows to sample a broad temperature range
by replacing the Boltzmann weight by an a priori unknown weight function that
is iteratively adjusted in order to yield a flat histogram.~\cite{JankeHist} To
this end, we first set an energy range obtained from exemplary parallel
tempering simulations of small systems. The weight function $W(E)$ is defined
on a discretized energy space with 1000 bins in the selected range. Between
consecutive iterations the weight function is updated by dividing each entry by
the amount of sampled data within the energy bin stored in a corresponding
histogram.  In order to achieve sufficient statistics for large systems with either
a large number of total monomers or a large sphere, we employ parallel
multicanonical simulations with up to 256 cores.~\cite{Zierenberg2013} This
parallelization efficiently distributes the required amount of statistics for
the weight iteration and speeds up the final data production linearly.  

Updates of the system are randomly drawn from a set of moves including
single-bead displacement, bond rotation, polymer translation as well as inter-
and intra-polymer rearrangement (double-bridging) moves. In order to increase
efficiency of the multicanonical method covering a broad energy range, we employ
energy-dependent update ranges. These are implemented such that detailed balance
is fulfilled.~\cite{Schnabel2011Adv}  Increasing the radius of the confining
sphere leads to a fast increase of conformational entropy. Thus, the maximal
translation step is coupled to the radius of the spherical confinement in order
to move polymers across larger distances when the system becomes more dilute.
In order to cope with the immense entropy gain properly, we also coupled the
number of sweeps per iteration and measurement linearly to the radius of the
sphere.

The data from the final equilibrium production run is afterwards reweighted to
yield canonical statistics in the desired temperature range $T\in[0.6,3.0]$
where we apply time-series reweighting ($\Delta T=0.1$ steps) including error
analysis and histogram reweighting ($\Delta T=0.005$ steps) for the connecting
lines in the canonical plots.~\cite{Reweighting} The aggregation transition
temperature is calculated by computing the second derivative of the total
energy and locating the zero crossing with an iterative time-series
reweighting. Errors are obtained from repeating this procedure in the framework
of jackknife error analysis. To this end, we combine all but one of the (up to
256) independent time series, calculating a set of highly correlated estimators
of the transition temperature. The jackknife error analysis now takes this
trivial correlation into account and provides an unbiased error estimate. This
is of advantage because the heat capacity is known to yield biased estimators,
which we may encounter for small data subsets.

\section{Results}
\label{sec:results}


\subsection{Canonical picture}

\begin{figure}
  \includegraphics[]{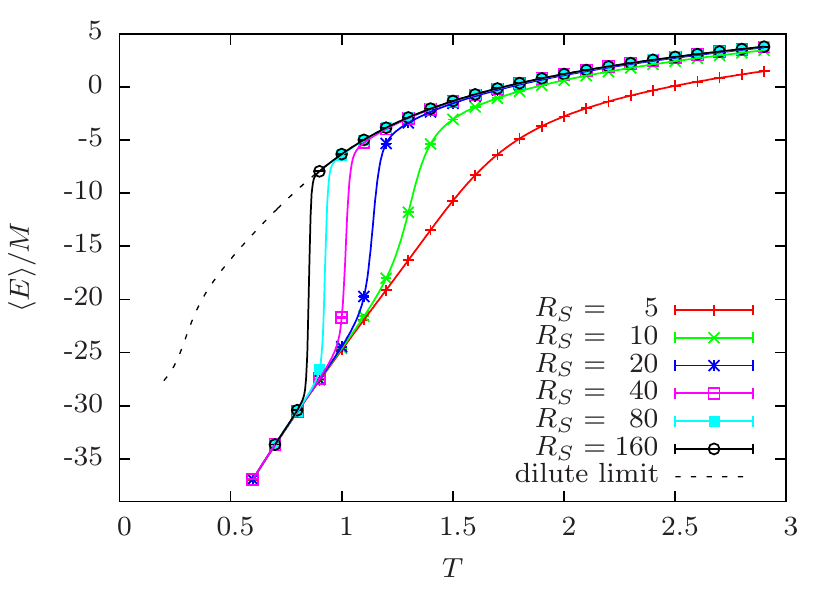}
  \includegraphics[]{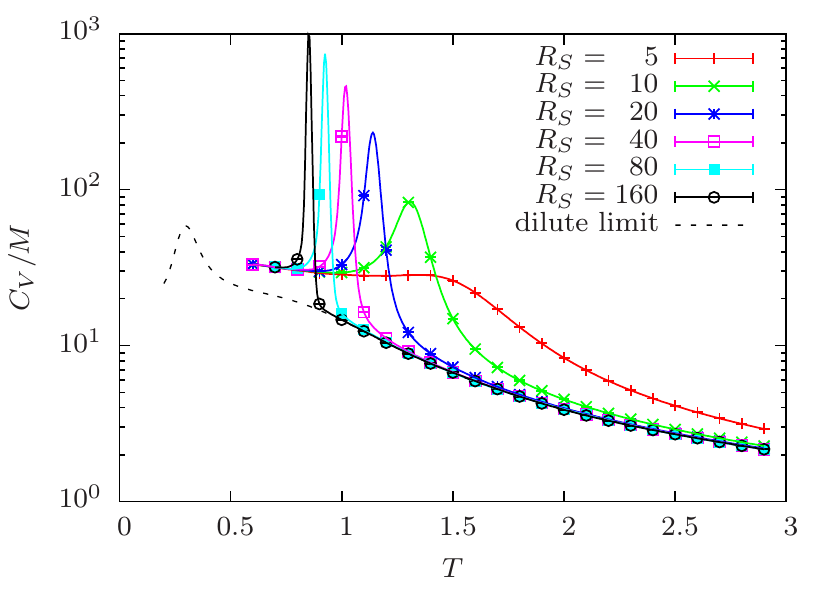}
  \caption{
    Normalized energy (top) and specific heat (bottom) for $M=8$ flexible polymers
    of length $N=13$ in spherical confinement. With increasing radius $R_S$, the
    density decreases and the aggregation transition shifts to lower
    temperatures. Notice, that the principle behavior at large temperature
    resembles the single polymer (dilute limit) behavior and that the dependence
    of the amorphous aggregate, at low temperature, follows the characteristics
    at larger density.
  }
  \label{fig:Canonical-Energy-Flexible}
\end{figure}

\begin{figure}
  \includegraphics[]{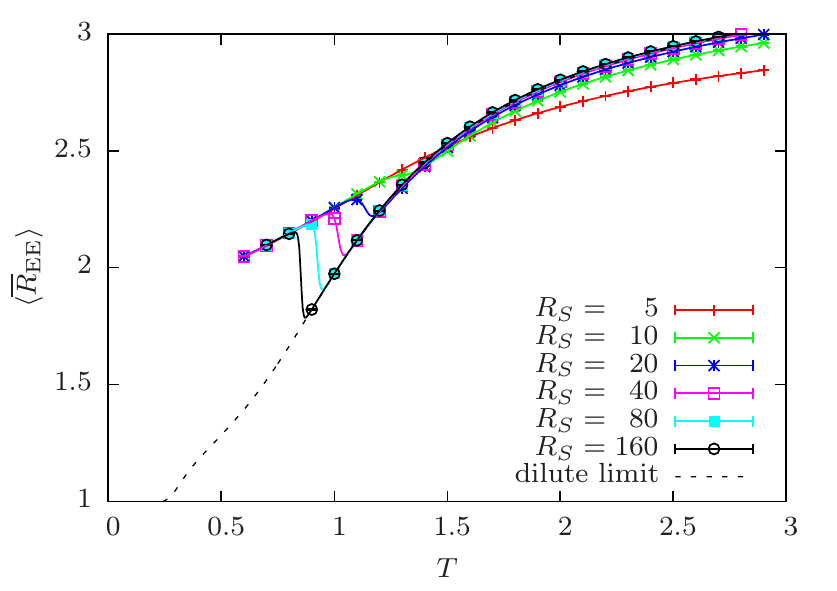}
  \includegraphics[]{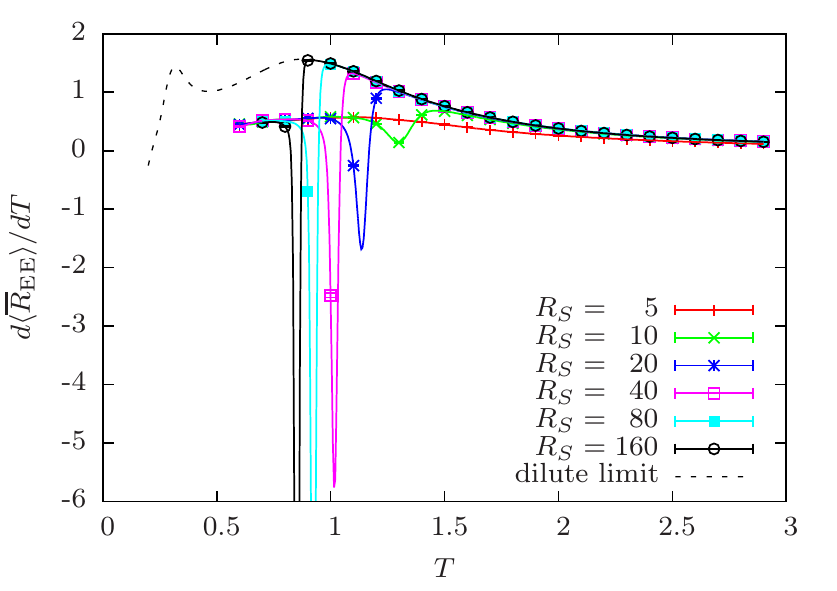}
  \caption{
    Average end-to-end distance {per polymer} (top) and its
    temperature derivative (bottom) for $M=8$ flexible polymers of length $N=13$
    in spherical confinement of size $R_S$. 
  }
  \label{fig:Canonical-EndToEnd-Flexible}
\end{figure}

Homogeneous aggregation describes the crossover from a separated phase of
individual polymers to an aggregated phase in which {a fraction
of} polymers will be condensed into a single {macroscopic}
object. In a strict sense, the notion of phases only applies to infinite
systems, whereas polymer systems are argued to be finite because the length of
a polymer is finite by nature.  However, we adapt the notion of phases to the
limit {of infinitely many polymers,} $M\rightarrow\infty$, at
fixed polymer density $\rho=M/V$ and fixed polymer length $N$, where
$V=\frac{4\pi}{3}R_S^3$ {is the volume of the confining sphere.
Then} the notion of phases can be regained in the usual way. In this case, we
consider the polymer length to be a system property just like the interaction
parameters. It may be expected that above the aggregation transition the
polymers will behave as in the dilute limit, exploring the conformational space
independently. Fixing the number of polymers and increasing the radius of the
confining sphere reduces the density and eventually leads to the dilute limit
of isolated polymer chains for $R_S\rightarrow\infty$.  In order to compare to
the dilute limit, we consider the system energy normalized to a single polymer
$E/M$ together with its thermal derivative, the specific heat per polymer
$C_V/M=\beta^2\left(\langle E^2\rangle - \langle E \rangle^2\right)/M$, where
$\beta=1/T$ (in units where $k_B=1$). In addition, we measure the average
end-to-end distance per polymer as the sum over distances between the first and
the last monomers each: $\overline{R}_{\rm EE}=\frac{1}{M}\sum_{i=1}^M
R_{\mathrm{EE},i}$, where $R_{\mathrm{EE},i}$ is the end-to-end distance of a
single polymer. For all observables, $\langle...\rangle$ denotes the thermal
average.

Figures \ref{fig:Canonical-Energy-Flexible} and
\ref{fig:Canonical-EndToEnd-Flexible} show the average energy and average
end-to-end distance with their thermal derivatives for 8 flexible polymers of
length $N=13$ for various sizes of spherical confinement. {The
lines are results from reweighting the raw data of the final multicanonical
production run. We present data points with error bars from an extensive
jackknife error analysis~\cite{Efron1982} at equidistant temperatures only.
The curve labeled ``dilute limit'' is obtained from a separate multicanonical
simulation of a single polymer.} Note that above the aggregation transition down
to the point of structural rearrangement into a single
{macroscopic} object, the polymers follow on average the behavior
of {the} isolated chain of length $N=13$ (dilute limit). This is
only hindered for small spheres, in this example $R_S=5$, where the separated
phase cannot be achieved because the cavity is of the order of the aggregate.
{This can be understood in terms of the overlap threshold
$\Phi^*$ of polymer solutions:~\cite{deGennes} If the volume fraction
$\Phi=NM\left(\frac{r_0}{2}\right)^3/R_S^3$ of a multi-polymer system is much
smaller than the intrinsic volume fraction of a single random coil,
\begin{equation}
  \Phi^*\simeq \frac{N\left(\frac{r_0}{2}\right)^3}{R_{\rm EE}^3}\simeq N^{1-3\nu}\approx
  N^{-0.76},
\end{equation}
where $R_{\rm EE}\simeq r_0N^\nu$ is the end-to-end distance of a self-avoiding walk
with $\nu\approx 0.588$, each polymer may be considered
independent and the system is dilute. A volume fraction of the order of this single
Gaussian-coil threshold, however, describes the onset of the semi-dilute region. Thus, we
look for the point where $\Phi = \Phi^*$. Solving this for the radius $R_S$ of the spherical
confinement, one sees that multi-polymer solutions may be considered dilute for radii
sufficiently larger than 
\begin{equation}
  R_S^c \simeq r_0M^{1/3}N^{\nu}.
\end{equation}
The crossover to the semi-dilute regime occurs around $R_S\approx R_S^c$. For
$N=13$, $R_S^c\approx 6.3$ which is consistent with the deviations we observe
for $R_S=5$ and to a lesser extent also for $R_S=10$.
}
Below the aggregation transition {in the dilute regime}, the
canonical observables coincide again because then the spherical confinement has
almost no effect on the structural properties of the aggregate.

This implies that the individual flexible polymers each follow the collapse
transition of the dilute limit down to the temperature where aggregation
suddenly sets in and becomes the major physical process determining the system's
equilibrium properties. For systems with equal inter- and intra-polymer
interactions, it has been noticed that collapse and aggregation are not separate
processes but that aggregation dominates.~\cite{JunghansPolymer} This is best
seen in Fig.~\ref{fig:Canonical-EndToEnd-Flexible} where the derivative of the
average end-to-end distance shows a broad peak around $T\sim0.9$ corresponding
to the collapse 
{of a single polymer (dilute limit). Sufficiently dilute systems
follow this behavior down to the point of aggregation.}
{ At first sight surprising is the increase of the average
end-to-end distance $\overline{R}_{\rm EE}$ at the aggregation transition
(Fig.~\ref{fig:Canonical-EndToEnd-Flexible} top).} {It can be
argued that} within the aggregate the average end-to-end distance is larger than
for the single collapsed polymer because the amorphous aggregates are highly
entangled{, forming a macroscopic} spherical object rather than
patching collapsed polymers together. The dominance of the aggregation
transition can be understood by its discontinuous nature opposed to the
{continuous} collapse transition. This discontinuous nature
{follows from a strong structural variation and can be seen as a
sharp jump of, e.g.,} the end-to-end distance.
{The size of the change increases } with decreasing density, see
Figs.~\ref{fig:Canonical-Energy-Flexible} and \ref{fig:Canonical-EndToEnd-Flexible}.  
{Moreover, at sufficiently small densities the polymers are
expected to aggregate at those temperatures where single polymers
assume globule conformations}. However, the energetic arguments in
Ref.~\onlinecite{JunghansPolymer} and the presented data suggest that even then
the collapsed conformations should unfold in order to form entangled aggregates
as equilibrium conformation. 

\begin{figure}
  \includegraphics[]{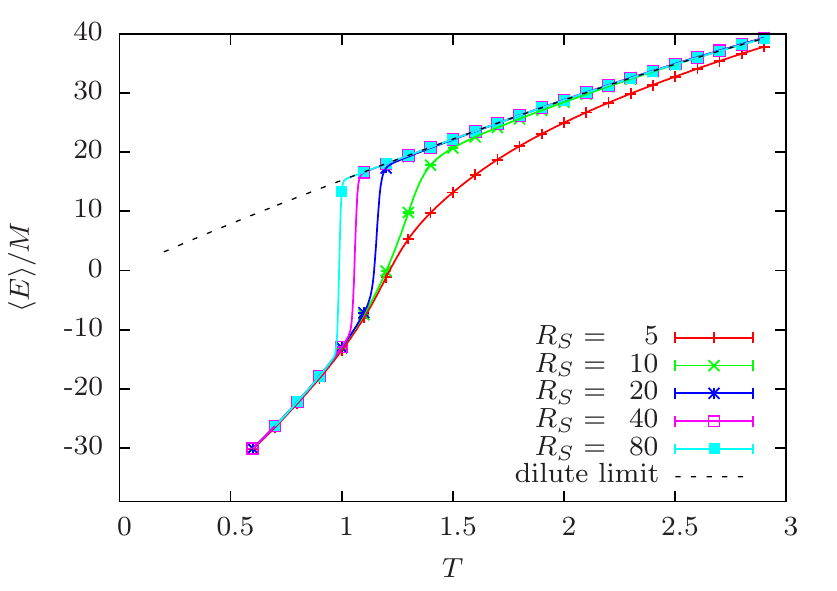}
  \includegraphics[]{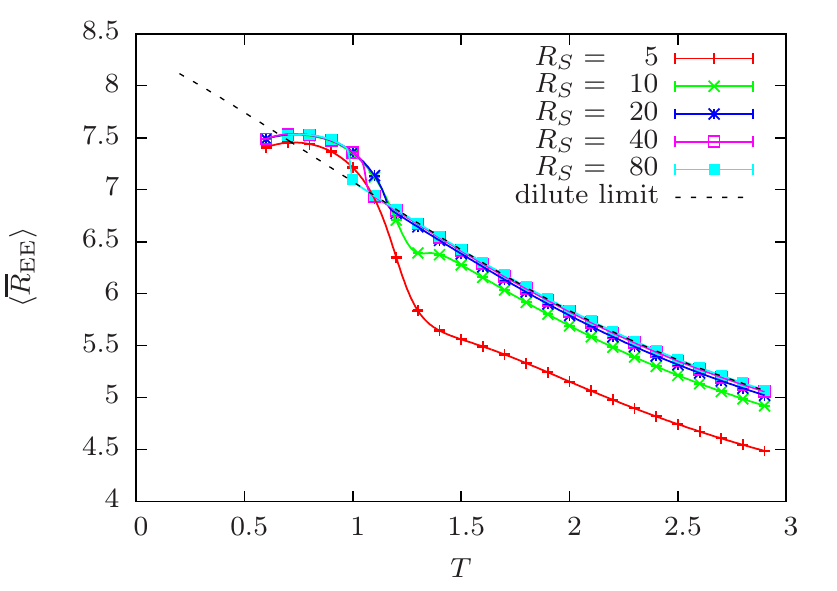}
  \caption{
    Normalized energy (top) and average end-to-end distance {per
    polymer} (bottom) for $M=8$ rather stiff polymers ($\kappa=9$) of length
    $N=13$ in a sphere of radius $R_S$.
  }
  \label{fig:Canonical-Stiff}
\end{figure}

The same observation of coinciding dilute and aggregated phases may be made for
rather stiff polymers. For eight polymers of length $N=13$ it was
shown~\cite{Zierenberg2014} that a bending stiffness with $\kappa=9$ leads to
polymer bundles in the aggregated phase. In this regime, the dilute limit of a
single polymer does not show a collapse transition, on the contrary the
individual polymers are driven to elongate in order to minimize curvature, see
Fig.~\ref{fig:Canonical-Stiff}. As in the case of flexible polymers, the
aggregation transition occurs only in a very narrow temperature range and
changes the average behavior from that of individual polymers to dense polymers
which in this case form polymer bundles. Interesting is that the formation of
polymer bundles in the sampled density range leads to an initial increase in the
average elongation right below the transition. However, a further reduction of
the temperature causes the system to form twisted bundles in which the average
end-to-end distance gets reduced again.~\cite{Zierenberg2014} It may be expected
that at even lower densities and consequently lower aggregation temperatures the
bundle formation may directly lead to twisted bundles.

As mentioned before, the equilibrium aggregation transition of homopolymers
leads to a single macroscopic aggregate instead of multiple smaller aggregates.
This can be observed in the total radius of gyration (see
Fig.~\ref{fig:ScalingRadGyr} with a detailed discussion in Sec.~\ref{sec:fss})
which drops to the scale of a single polymer of length $NM$. Despite the
non-vanishing probability to form metastable states with several aggregates,
these conformations seem to not affect the equilibrium properties significantly
away from the transition point.

%
%
\subsection{Microcanonical picture}
\label{sec:microcanonical}

The spherical confinement provides a safe and controllable base to study the
effect of density on the aggregation transition. In principle, a periodic box
would also allow to study this effect. However, the possibility that the
aggregate may interact with itself across the boundaries introduces systematic
errors, which are hard to tackle as the probability of self-interactions rises
with increasing density. This effect gets excluded with the spherical
confinement, while introducing at most an effective repulsion from the steric wall.  

Here, we consider the number of monomers per polymer $N$ to be a system
property that describes the extension of a polymeric object. This is a valid
assumption in the dilute limit and in the limit of many polymers, where we
focus on the former. In order to show the generality of our results, we
consider parameters \mbox{$N=\{13,20,27\}$} which will show the same
qualitative behavior. A more detailed discussion of the influence of $N$ will
be given in Sec.~\ref{sec:fss}.

In order to quantify the effect of density on the aggregation transition, we
consider the microcanonical ensemble and start out with the Gibbs construction.
The microcanonical entropy is given by the logarithm of the total number
$\Omega$ of configurations with a given energy: $S(E)=\ln\Omega(E)$.
Moreover, the microcanonical inverse temperature is defined as the local slope
of this microcanonical entropy, or in different words its derivative with
respect to $E$. Expecting a phase coexistence, there should exist an inverse
temperature as local slope to two energy states, equivalent to considering an
energy distribution with two peaks in the canonical ensemble.  This allows us to
estimate the inverse aggregation temperature as the slope of the hull connecting
the microcanonical entropy of the aggregated and the separated
phase.~\cite{JankeMicro,JunghansPeptide,JunghansPolymer} Hence,
\begin{equation}
  \beta_{\rm agg} = \frac{S(E_{\rm sep}) - S(E_{\rm agg})}{\Delta E}.
\end{equation}

\begin{figure}
  \includegraphics[]{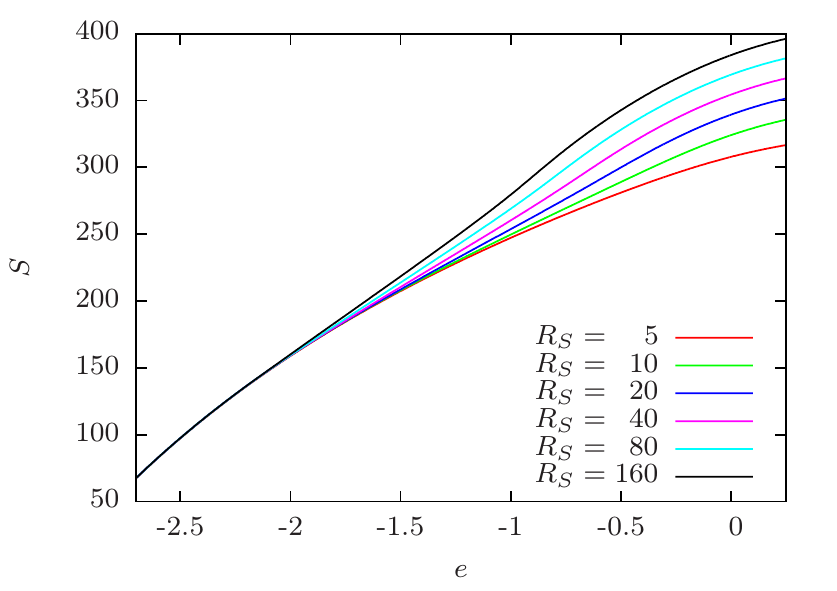}
  \caption{
    The microcanonical entropy $S$ (up to an additive constant) as the
    logarithmic density of states $\ln\hat\Omega$ for $M=8$ polymers ($N=13$),
    obtained from a microcanonical analysis of the multicanonical data. 
  }
  \label{fig:Entropy}
\end{figure}
Now, consider a spherical confinement of radius $R_S$. Figure~\ref{fig:Entropy}
shows the microcanonical entropy for $8$ polymers of length $N=13$ obtained as
the logarithm of the estimated density of states $\hat{\Omega}(E)$. An estimate
of the density of states comes directly from the multicanonical method within
the selected energy range when dividing the final histogram by the weight
function $\hat{\Omega}(E)=H(E)/W(E)$. The number of states in the aggregated
phase will barely be influenced by the confinement, compared to the ensemble of
polymers in the fluctuating phase that behaves more like a gas. Therefore, we
assume for the separated phase that the number of states will be proportional to
$V^M$ as in the case of an ideal gas,
\begin{equation}
  S(E_{\rm sep}) \sim \ln\left[ \left(\frac{4\pi}{3} R_S^3\right)^M \right]\propto
  M\ln R_S,
\end{equation}
thus dominating over $S(E_{\rm agg})$. Assuming that the latent heat will be
almost constant with respect to $R_S$ for fixed $(M,N)$, we may write $\Delta
E=M\Delta e$. This leads to an aggregation temperature depending on the
logarithm of the radius of the confining sphere:
\begin{equation}
  \beta_{\rm agg}(R_S) \sim \frac{S(E_{\rm sep})}{\Delta E}  \sim \ln R_S + \mathrm{const}.
  \label{eq:ansatz_R}
\end{equation}
This may be rewritten in terms of the density $\rho$ as well $\beta_{\rm agg}=a_1\ln
\rho + a_2$, which has been observed recently, e.g., for two lattice
proteins~\cite{Ni2013} and polymer adsorption~\cite{moeddel2010}.

\begin{figure}
  \includegraphics[]{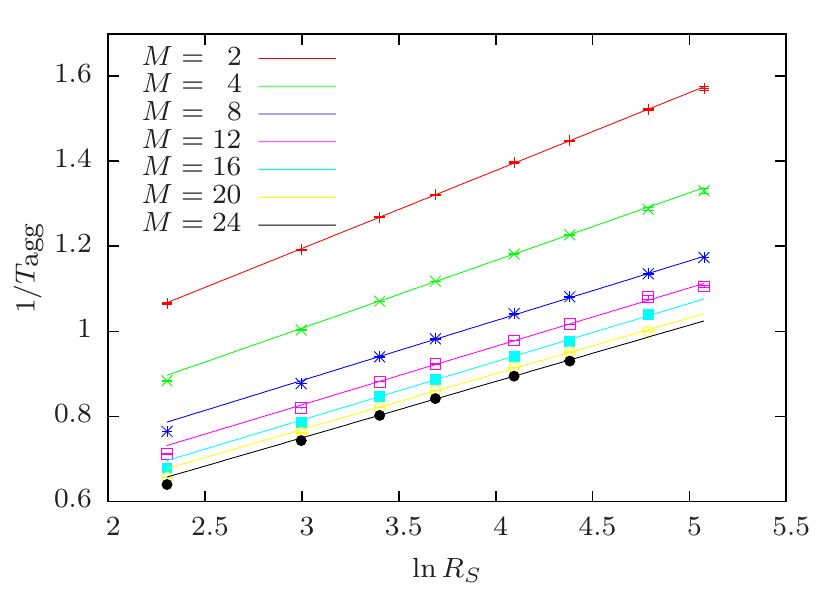}
  \caption{
    Scaling of the inverse aggregation temperature with the logarithm of the system size
    for $N=13$. For more detail see Table \ref{tab:N013}.
  }
  \label{fig:Scaling_R}
\end{figure}
Tables \ref{tab:N013}, \ref{tab:N020}, and \ref{tab:N027} present aggregation
transition temperatures, obtained from the peak location of the specific heat
$C_V/M$, for a wide range of polymer sizes \mbox{$N=\{13,20,27\}$} and polymer
numbers \mbox{$M=\{2,4,8,12,16,20,24\}$} as far as possible.
Figure~\ref{fig:Scaling_R} shows the inverse aggregation temperature as a
function of $\ln R_S$ for the example $N=13$. It can be seen that the expected
scaling (\ref{eq:ansatz_R}) is clearly confirmed by the data and that with
increasing $M$ the slopes of the fit become more similar.
\renewcommand{\arraystretch}{1.0}
\begin{table}[h]
  \caption{
    Aggregation temperatures obtained from the peak location of the specific heat for
    $N=13$ and $R_S\ge20$. In addition, we present the fit results from
    $\beta=a\ln R_S+b$ for $R_S\ge 30$.
  }
  \label{tab:N013}
  \baselineskip=10pt
  \begin{tabularx}{\columnwidth}{r c r c l c}
  \hline
  $M$ & $\quad$ & $R_S$ & $\qquad$ & $T_{\rm agg}$ & fit parameters \\
  \hline
  2 &  & 20 &  &   0.8390(6)  &\\
  2 &  & 30 &  &   0.7884(6)  &\\
  2 &  & 40 &  &   0.7569(5)  &\\
  2 &  & 60 &  &   0.7161(14) &  0.1831(12)~$\ln R_S$~+~0.646(5)\\
  2 &  & 80 &  &   0.6903(6)  &\\
  2 &  & 120 &  &   0.6572(11)  &\\
  2 &  & 160 &  &   0.6367(18)  &\\
  \hline4 &  & 20 &  &   0.9967(12)  &\\
  4 &  & 30 &  &   0.9337(4)  &\\
  4 &  & 40 &  &   0.8944(4)  &\\
  4 &  & 60 &  &   0.8462(9) &  0.1588(9)~$\ln R_S$~+~0.531(4)\\
  4 &  & 80 &  &   0.8151(7)  &\\
  4 &  & 120 &  &   0.7769(20)  &\\
  4 &  & 160 &  &   0.752(4)  &\\
  \hline8 &  & 20 &  &   1.1399(4)  &\\
  8 &  & 30 &  &   1.06385(26)  &\\
  8 &  & 40 &  &   1.0179(4)  &\\
  8 &  & 60 &  &   0.9601(6) &  0.14063(23)~$\ln R_S$~+~0.4628(10)\\
  8 &  & 80 &  &   0.92494(22)  &\\
  8 &  & 120 &  &   0.8807(9)  &\\
  8 &  & 160 &  &   0.85221(27)  &\\
  \hline12 &  & 20 &  &   1.2184(4)  &\\
  12 &  & 30 &  &   1.1338(5)  &\\
  12 &  & 40 &  &   1.0832(7)  &\\
  12 &  & 60 &  &   1.0216(7) &  0.1367(6)~$\ln R_S$~+~0.4177(20)\\
  12 &  & 80 &  &   0.9834(5)  &\\
  12 &  & 120 &  &   0.926(4)  &\\
  12 &  & 160 &  &   0.9044(13)  &\\
  \hline16 &  & 20 &  &   1.2717(5)  &\\
  16 &  & 30 &  &   1.1813(5)  &\\
  16 &  & 40 &  &   1.1278(6)  &\\
  16 &  & 60 &  &   1.0626(7) &  0.1363(8)~$\ln R_S$~+~0.3830(27)\\
  16 &  & 80 &  &   1.0244(13)  &\\
  16 &  & 120 &  &   0.9611(18)  &\\
  \hline20 &  & 20 &  &   1.3126(11)  &\\
  20 &  & 30 &  &   1.21752(26)  &\\
  20 &  & 40 &  &   1.1611(4)  &\\
  20 &  & 60 &  &   1.0929(8) &  0.1315(5)~$\ln R_S$~+~0.3748(15)\\
  20 &  & 80 &  &   1.0504(9)  &\\
  20 &  & 120 &  &   0.9998(7)  &\\
  \hline24 &  & 20 &  &   1.3451(8)  &\\
  24 &  & 30 &  &   1.2459(8)  &\\
  24 &  & 40 &  &   1.1873(8) &  0.1306(8)~$\ln R_S$~+~0.3592(29)\\
  24 &  & 60 &  &   1.1177(6)  &\\
  24 &  & 80 &  &   1.0752(9)  &\\
  \hline
  \end{tabularx}
\end{table}
\begin{table}[h]
  \caption{
    Same as Table~\ref{tab:N013} for $N=20$
  }
  \label{tab:N020}
  \begin{tabularx}{\columnwidth}{r c r c l c}
  \hline
  $M$ & $\quad$ & $R_S$ & $\qquad$ & $T_{\rm agg}$ & fit parameters \\
  \hline
  2 &  & 20 &  &   1.0373(10)  &\\
  2 &  & 30 &  &   0.9793(7)  &\\
  2 &  & 40 &  &   0.9430(15)  &\\
  2 &  & 60 &  &   0.8944(10) &  0.1371(11)~$\ln R_S$~+~0.555(4)\\
  2 &  & 80 &  &   0.8649(11)  &\\
  2 &  & 120 &  &   0.8266(12)  &\\
  2 &  & 160 &  &   0.801(4)  &\\
  \hline4 &  & 20 &  &   1.2034(10)  &\\
  4 &  & 30 &  &   1.1331(10)  &\\
  4 &  & 40 &  &   1.0884(10)  &\\
  4 &  & 60 &  &   1.0342(6) &  0.1200(10)~$\ln R_S$~+~0.475(4)\\
  4 &  & 80 &  &   0.9995(9)  &\\
  4 &  & 120 &  &   0.9549(23)  &\\
  \hline8 &  & 20 &  &   1.3513(7)  &\\
  8 &  & 30 &  &   1.2685(5)  &\\
  8 &  & 40 &  &   1.2179(12)  &\\
  8 &  & 60 &  &   1.1547(18) &  0.1083(4)~$\ln R_S$~+~0.4202(15)\\
  8 &  & 80 &  &   1.1156(11)  &\\
  8 &  & 120 &  &   1.0659(6)  &\\
  \hline12 &  & 20 &  &   1.4331(6)  &\\
  12 &  & 30 &  &   1.3408(10)  &\\
  12 &  & 40 &  &   1.2861(7)  &\\
  12 &  & 60 &  &   1.2185(27) &  0.1037(6)~$\ln R_S$~+~0.3943(22)\\
  12 &  & 80 &  &   1.1772(16)  &\\
  12 &  & 120 &  &   1.1163(19)  &\\
  12 &  & 160 &  &   1.0894(12)  &\\
  \hline16 &  & 20 &  &   1.4876(4)  &\\
  16 &  & 30 &  &   1.3903(7)  &\\
  16 &  & 40 &  &   1.3319(5) &  0.1058(6)~$\ln R_S$~+~0.3602(21)\\
  16 &  & 60 &  &   1.2601(5)  &\\
  16 &  & 80 &  &   1.2182(11)  &\\
  \hline20 &  & 20 &  &   1.5296(7)  &\\
  20 &  & 30 &  &   1.4267(5)  &\\
  20 &  & 40 &  &   1.3661(7) &  0.1034(5)~$\ln R_S$~+~0.3495(17)\\
  20 &  & 60 &  &   1.2928(10)  &\\
  20 &  & 80 &  &   1.2466(6)  &\\
  \hline
  \end{tabularx}
\end{table}
\begin{table}[h]
  \caption{
    Same as Table~\ref{tab:N013} for $N=27$
  }
  \label{tab:N027}
  \begin{tabularx}{\columnwidth}{r c r c l c}
  \hline
  $M$ & $\quad$ & $R_S$ & $\qquad$ & $T_{\rm agg}$ & fit parameters \\
  \hline
  2 &  & 20 &  &   1.1786(15)  &\\
  2 &  & 30 &  &   1.1172(21)  &\\
  2 &  & 40 &  &   1.0785(13)  &\\
  2 &  & 60 &  &   1.0276(12) &  0.1114(10)~$\ln R_S$~+~0.517(5)\\
  2 &  & 80 &  &   0.9942(14)  &\\
  2 &  & 120 &  &   0.9524(9)  &\\
  2 &  & 160 &  &   0.9249(13)  &\\
  \hline4 &  & 20 &  &   1.3489(7)  &\\
  4 &  & 30 &  &   1.2745(14)  &\\
  4 &  & 40 &  &   1.2279(18)  &\\
  4 &  & 60 &  &   1.1701(16) &  0.0985(10)~$\ln R_S$~+~0.450(4)\\
  4 &  & 80 &  &   1.1343(12)  &\\
  4 &  & 120 &  &   1.084(4)  &\\
  4 &  & 160 &  &   1.0547(28)  &\\
  \hline8 &  & 20 &  &   1.4981(10)  &\\
  8 &  & 30 &  &   1.4115(8)  &\\
  8 &  & 40 &  &   1.3574(9) &  0.0925(7)~$\ln R_S$~+~0.3945(24)\\
  8 &  & 60 &  &   1.2935(10)  &\\
  8 &  & 80 &  &   1.2515(10)  &\\
  \hline12 &  & 20 &  &   1.5783(16)  &\\
  12 &  & 30 &  &   1.4847(9)  &\\
  12 &  & 40 &  &   1.4268(10) &  0.0897(6)~$\ln R_S$~+~0.3691(20)\\
  12 &  & 60 &  &   1.3564(19)  &\\
  12 &  & 80 &  &   1.3128(7)  &\\
  \hline16 &  & 20 &  &   1.6338(7)  &\\
  16 &  & 30 &  &   1.5332(6)  &\\
  16 &  & 40 &  &   1.47264(28) &  0.0907(8)~$\ln R_S$~+~0.3444(27)\\
  16 &  & 80 &  &   1.3533(16)  &\\
  \hline
  \end{tabularx}
\end{table}
\renewcommand{\arraystretch}{1.0}

\subsection{Finite-size effects at polymer aggregation}
\label{sec:fss}

\begin{figure}
  \includegraphics[]{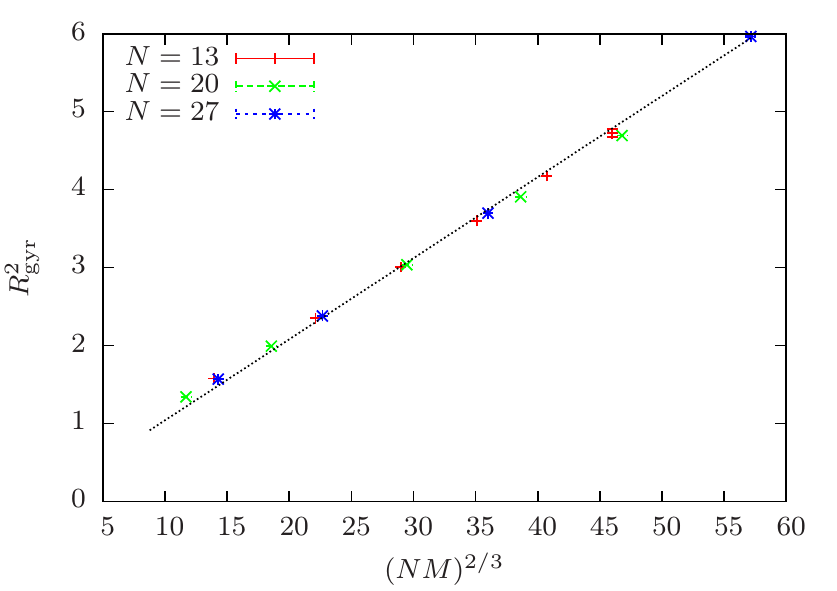}
  \caption{
    Data collapse of the squared radius of gyration as a function of
    $(NM)^{2/3}$.  The polymer systems are in spheres of size $R_S=30$ with
    $N=\{13,20,24\}$ at $T=0.7$.
  }
  \label{fig:ScalingRadGyr}
\end{figure}

The effect of stiffness on aggregation of a few polymers in bulk has been
investigated recently.~\cite{Zierenberg2014} It was shown that the polymers
form amorphous aggregates in the case of flexible polymers and correlated
polymer bundles in the stiff polymer limit. Here, we focus on the flexible
limit, where the amorphous aggregate may be argued to behave similarly to a single
polymer of length $NM$. Moreover, in this limit it should result in a spherical
aggregate with radius $R\sim\left(NM\right)^{1/3}$ as predicted for the
collapsed isolated polymer. This should be observable in the total squared
radius of gyration
\begin{equation}
  R_{\rm gyr}^2=\frac{1}{NM} \sum \left(\vect{r}_i-\vect{r}_{\rm cm}\right)^2,
\end{equation}
where $\vect{r}_{\rm cm}$ is the center of mass vector of the total system.  For
small spheres ($R_S=30$) the polymers form stable aggregates at sufficiently
high temperatures. Furthermore, the canonical results demonstrate that the
amorphous aggregates behave consistently, despite different radii.
Figure~\ref{fig:ScalingRadGyr} shows the squared radius of gyration $R_{\rm
gyr}^2$ of the total system versus the expected scaling function of the total
number of monomers $f(NM)=(NM)^{2/3}$.  Because our focus is to sample the
aggregation transition temperature, we consider a reduced energy range going to
sufficiently small energies.  However, this does not allow to extrapolate to
temperatures far below the aggregation transition temperature. In order to
compare the scaling of the squared radius of gyration, we need a fixed
temperature that is below the aggregation transition temperature but which is
still within the sampled range of our simulations. Both boundaries vary with the
length and number of polymers.  This leads to a relatively small sample size in
Fig.~\ref{fig:ScalingRadGyr} because $T=0.7$ is so large that the smallest
systems are not yet in the aggregated state but too small for the large systems
to be sampled with the chosen energy range.  Nonetheless, it shows the
qualitative data collapse that we expected for finite polymer systems.

In general, the idea of finite-size scaling is to use small systems in order to
make predictions about a system of infinite size and to describe how certain
physical properties are approached with increasing system size.  In spin systems
for example, one may consider the deviation of the transition temperature
between the ordered and disordered phases from the value of the infinite system
$L\rightarrow\infty$. For polymer aggregation, however, this limit becomes
nontrivial. Single polymer studies for example consider finite-size scaling for
$N\rightarrow\infty$.  In systems with more than one polymer, we have to
introduce some sort of boundary conditions and define a density. For the limit
of infinitely long polymers, we were not able to think of a density with a
well-defined scaling behavior, not changing local system properties. 

Thus, we fix $N$ (considering three realizations) and investigate
deviations for different $M,R_S$ at polymer density $\rho=M/V$ constant. This
choice does not change the average local system in the infinite system limit
$M\rightarrow\infty$. Although our system sizes are rather small for a
quantitatively accurate finite-size scaling, we attempt to draw conclusions about
the finite-size effects in small systems.

{For systems with a small number of polymers}, it is a reasonable
assumption that the aggregation transition separates a gas-like phase (see
arguments in Sec.~\ref{sec:microcanonical}) from a homogeneous aggregate and
{approximate} void space. For {a larger number of
polymers in the dilute regime}, there are good arguments for a mixed phase of a
macroscopic aggregate and a polymer gas phase in analogy to particle
condensation.~\cite{Nussbaumer2008,Zierenberg2014JPCS} {From our
experience with particle systems, we expect to observe even for moderately large
system sizes a large fraction of polymers going into the equilibrium aggregate.}
In any case, the finite aggregate forms a surface such that the system may be
compared to ``single-domain'' non-periodic boundary conditions.  The finite-size
corrections then are of the order $1/L$, where $L$ is the characteristic system
length (for details see also Refs.~\onlinecite{Privman1990,Borgs1995}). This may
be interpreted as competing contributions from the system volume ($\sim L^d$)
and system surface ($\sim L^{d-1}$). We make use of this observation and
consider the linear extension of the aggregate $R_{\rm gyr}$ as characteristic
length scale of the system. As we showed in Fig.~\ref{fig:ScalingRadGyr}, for
flexible polymers the aggregates are spherical and for small systems
{they include a large fraction of the polymers such that} $R_{\rm
gyr}\sim M^{1/3}$, where we omitted $N$ which is considered as a fixed system
parameter.  Thus, our ansatz for monodisperse polymers is
\begin{equation}
  T_{\rm agg}(M,\rho) \propto \left(1+s(\rho)M^{-1/3}+\mathcal{O}\left(M^{-2/3}\right)\right),
  \label{eq:fss_ansatz}
\end{equation}
where $s(\rho)$ determines the size of the leading correction and may depend on
the density.

\begin{figure}
  \includegraphics[]{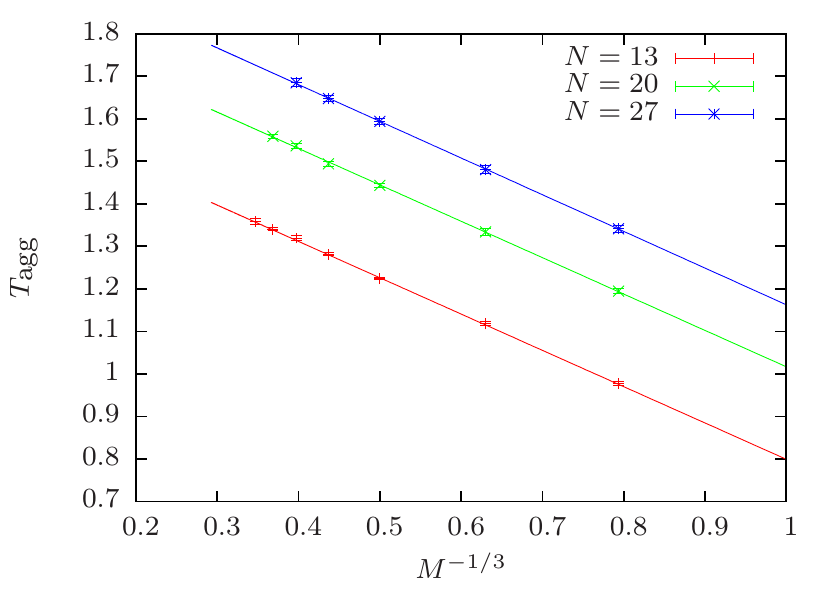}
  \caption{
    Fit of the finite-size effects for all $N$ at fixed density $\rho=10^{-3}$.
    Error bars are obtained by error propagation and neglect possible systematic
    deviations due to higher-order corrections.
  }
  \label{fig:fss_T_M}
\end{figure}
An example of the ansatz is shown in Fig.~\ref{fig:fss_T_M} for the selected $N$
at fixed density $\rho=10^{-3}$. We fitted the data in Tables
\ref{tab:N013}-\ref{tab:N027} to the dependence (\ref{eq:ansatz_R}) and used
those fits to interpolate the inverse transition temperature to various dilute
polymer densities. Fitting the ansatz for the finite-size effects we obtained
the same qualitative results for different densities. As expected, the
extrapolated limit would again be depending on $\ln\rho$ (not shown here) and
moreover would most probably not be in the infinite system limit. The fit parameter
$s$ only slightly depends on the density. Thus, we assume $s$ for the remaining
part of the study to be constant. As we mentioned before, we consider $N$ as a
system parameter and thus it is not surprising that $s$ is $N$-dependent.

\begin{figure}
  \includegraphics[]{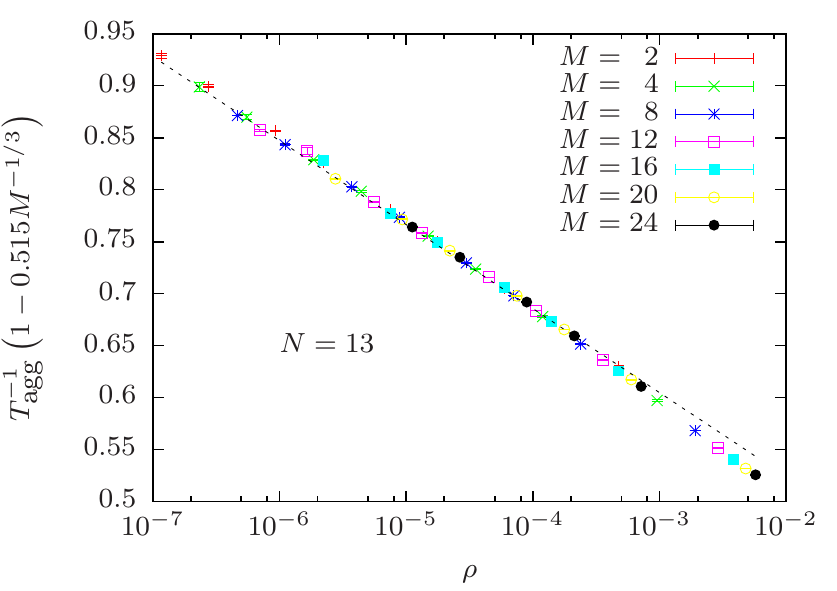}
  \includegraphics[]{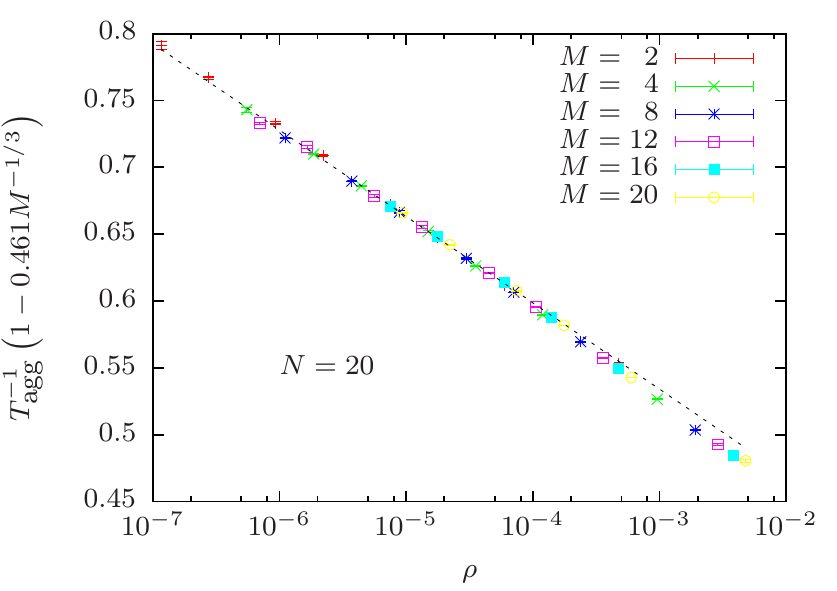}
  \includegraphics[]{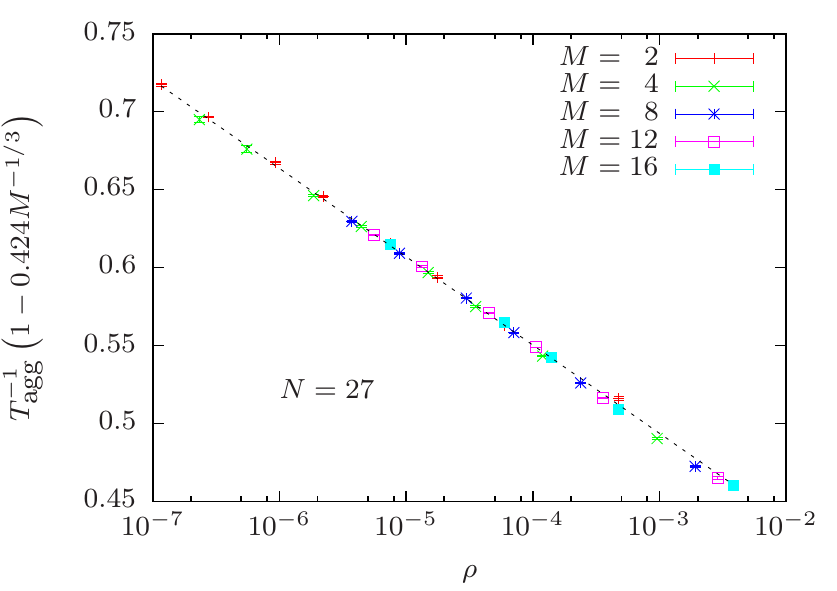}
  \caption{
    Rescaled inverse aggregation temperature versus density $\rho$ for all three
    polymer lengths $N=\{13,20,27\}$. The finite-size correction $s$ in Eq.
    (\ref{eq:fss_ansatz}) was assumed to be constant. For $N=13$ one may compare
    to Fig.~\ref{fig:Scaling_R}.
  }
  \label{fig:fss_collapse}
\end{figure}

We can combine the finite-size effects in order to rescale the inverse
aggregation temperature by multiplying it with the ansatz (\ref{eq:fss_ansatz}).
Considering the polymer density $\rho$ as the relevant parameter, we would then
expect that the data falls on a coinciding linear relation.
Figure~\ref{fig:fss_collapse} shows a reasonable data collapse for all simulated
polymer lengths, where we assumed $s$ constant with its value denoted on the
$y$-axes of the plots. The data collapses on a single line {is} consistent with the
developed density dependence and finite-size effects. Despite the $N$-dependent
scaling form, this shows that the entropic arguments for the density dependence
together with the energetic arguments for the finite-size effects are consistent
and allow a reasonably well description of the aggregation transition of a few
dilute monodisperse polymers in a spherical cavity.

\section{Conclusions}
\label{sec:conclusion}
Throughout our systematic investigation, we have demonstrated that the separated
phase of flexible and stiff polymers corresponds to the dilute limit for
sufficiently small densities. The aggregated phase itself is barely influenced
by the confinement. In the case of flexible polymers the aggregate may be
described as a spherical object, whose size scales like the collapsed state of a
single flexible polymer with the same total number of monomers. The spherical
confinement effects the location of the aggregation transition of dilute
polymers. For denser systems, the separated phase may be suppressed completely;
increasing the density even further the aggregate itself will be compressed by
the confinement, probably driving the system into its frozen state.

For the case of dilute polymers, we have presented entropic
(Sec.~\ref{sec:microcanonical}), geometric and energetic arguments
(Sec.~\ref{sec:fss}) that allow a description of the rescaled aggregation
temperature as a function of density for monodisperse flexible theta polymers.
The entropic considerations suggest a linear dependence of the inverse
aggregation temperature on the logarithm of the density. We deduced a
description of finite-size effects (\ref{eq:fss_ansatz}) that characterizes the
deviations among small systems reasonably well. The system sizes we investigated
are most likely too small to capture the behavior of infinite systems. However,
for experiments investigating small polymeric systems in confined geometries on
the nanoscale, these finite-size effects should be apparent.

The results presented are potentially relevant for experiments with dilute
polymer solutions, which investigate single polymer behavior. We showed that
even in the case of dilute solutions there exists an aggregation transition that
may be estimated by considering a few polymers in sufficiently dilute systems
and extrapolating to the dilute limit. In experiments, depending on the actual
free-energy barrier and the corresponding relaxation times, it would probably be
difficult to observe homogenous aggregation. In principle, estimates of this
barrier should be accessible to computer simulations using a proper finite-size
scaling ansatz together with sufficiently large system sizes. Our results
suggest that this may be done for rather dense systems, which would allow to reach
appropriate system sizes -- but that remains still to be done in a forthcoming
study.

\begin{acknowledgments}
The project was funded by the European Union and the Free State of Saxony.
Part of this work has been financially supported by
 the Collaborative Research Center SFB/TRR 102 (project B04), 
 the ESF Junior Research Group No.\ 241\,202 ``Function through
 Selforganization: Emergent Properties of Atomic and Molecular Aggregates'',
 the Leipzig Graduate School of Excellence GSC185 ``BuildMoNa'',
 and the Graduate College of the Deutsch-Franz\"osische Hochschule (DFH-UFA)
 under grant No.\ CDFA-02-07.
The computing time provided by the John von Neumann Institute for Computing
(NIC)  on the supercomputer JUROPA at J\"ulich Supercomputing Centre (JSC) under
grant \mbox{No. HLZ21}
is gratefully acknowledged.

\end{acknowledgments}

\bibliographystyle{model1-num-names}

\newpage

\end{document}